# Universal scaling of the Anomalous Hall voltage of magnetic layer with the adjacent conducting layer

Zhihao Yan[1,2], Qianbiao Liu[1], Zhengxiao Li[1,2], Lijun Zhu[1,2*]

1. *State Key Laboratory of Semiconductor Physics and Chip Technologies, Institute of Semiconductors, Chinese Academy of Sciences, Beijing 100083, China*

2. *Center of Materials Science and Optoelectronics Engineering, University of Chinese Academy of Sciences, Beijing 100049, China*

*ljzhu@semi.ac.cn

The anomalous Hall voltage of magnetic heterostructures plays a key role as the indicator for the magnetization state and its interplay with a variety of spintronic effects. In this letter, we report the observation, mechanism, and impact of the universal, dramatic scaling of the anomalous Hall voltage of magnetic heterostructures with the thickness and resistivity of the nonmagnetic conducting layers (e.g., normal metal) by combining transport experiments, analytical derivation, and finite-element analyses. We identify that the mechanism is the serial resistor effect of the magnetic and non-magnetic layers within the magnetic heterostructures and irrelevant to any Fermi surface variation or angular momentum injection from the nonmagnetic layer to the magnetic layer. We also show that the accuracy of the harmonic Hall voltage analyses of magnetic heterostructures are unaffected by the under-measuring of the anomalous Hall voltage and other transverse voltages due to the serial resistor effect. These findings provide crucial information for understanding a variety of spintronic phenomena involving the anomalous Hall and other transverse voltages.

*Introduction.* The anomalous Hall effect (AHE) of magnetic heterostructures has been playing a key role in spintronics. First of all, the anomalous Hall voltage ($V_{\rm AH}$) under a given electric field $E$, or equivalently the anomalous Hall resistance, has been widely utilized as the readout for magnetic fields [1-8], the magnetization states [11-17], magnetic domains [18,19], magnetic skyrmions [20-24], and interfacial spin-orbit coupling [25-27]. Furthermore, $V_{\rm AH}$ and its equivalent correlate the first and second harmonic Hall voltages ($V_{1\omega}$ and $V_{2\omega}$) to the dampinglike spin-orbit torque (SOT) in bilayers of normal metal (NM)/metallic ferromagnet [28-34], superlattices[35,36], magnetic single layers [39-40], synthetic antiferromagnet[41-45], antiferromagnet [46,47], van der Waals magnets [48,49], or magnetic oxides [50,51]. $V_{\rm AH}$ can also monitor the magnetization-dependent interfacial scattering of angular momentum and electrons in unusual magnetoresistance experiments [52,53] and indicate the tilting or switching of the magnetization driven by magnetic field [54-56], the Dzyaloshinskii–Moriya interaction [57-60], or spin currents [11-17, 54,55,61,62]. A recent study has also speculated that the AHE can be altered by spin and orbital currents [63]. Thus, it is of critical importance for a variety of spintronic experiments to accurately understand $V_{\rm AH}$ of magnetic heterostructures.

It has been well established that $V_{\rm AH}$ of a magnetic single layer originates from the band structure (intrinsic mechanism) and impurity scattering (side jump and skew scattering)[64-70]. In the absence of any adjacent conducting layer, $V_{\rm AH}$ of a magnetic single layer (we note as $V_{\rm AH,0}$) is principally given by

$$V_{\rm AH,0} = \rho_{\rm AH} E W/\rho_{\rm FM}, \quad (1)$$

where $\rho_{\rm AH}$, $W$, and $\rho_{\rm FM}$ are the anomalous Hall resistivity, the width of the Hall device, and the resistivity of the magnetic layer, respectively. However, it has remained a crucial open question as to whether and how $V_{\rm AH}$ of a magnetic layer within spintronic devices should be generally affected by the adjacent NMs.

In this letter, we perform a systematic study on seven different NM/magnet device series with varying thicknesses ($d$) and resistivities of the NM by combining transport experiments, analytical derivation, and finite-element analyses. We report the observation, mechanism, and impact of the universal scaling of the anomalous Hall voltage of NM/magnet heterostructures with $d$.

*Devices and methods.* For this study, we sputter-deposited six NM/magnet bilayer series, i.e., $Pt_{75}Au_{25}$ 4-8/Co 0.8, $Pt_{75}Pd_{25}$ 3-8/Co 0.6, Fe 2.5/Pt 1.4-5, $Fe_{50}Pt_{50}$ 2.5/Cu 0-10, Ti 1.7-6.9/$Fe_{60}Co_{20}B_{20}$ 2, Mn 2.5-9.1/$Fe_{55}Tb_{45}$ 8, and a trilayer series Pt 5/Co 2/Al 3-9.6. Here, the Pt [11,32,71], $Pt_{75}Au_{25}$ [72], and $Pt_{75}Pd_{25}$ [73] are chosen as the heavy metals with strong spin Hall effect, Cu [30], Ti [30,74-76], Mn [77], Al [78] are the representatives of light metals with negligible spin Hall effect. Fe, Co, $Fe_{60}Co_{20}B_{20}$, and $Fe_{50}Pt_{50}$ are representative metalic ferromagnets (FMs), and $Fe_{55}Tb_{45}$ is a typical ferrimagnet [40]. Each sample is protected by a MgO 2/$TaO_x$ 4 bilayer and fabricated into two-cross Hall bar devices with width ($W$) of 5 μm and length ($L$) of 60 μm (Fig. 1b). Finally, Ti 5/Pt 150 contacts were deposited for transport measurements. For each series, $V_{\rm AH}$ is measured utilizing a lock-in amplifier (SR860) under a fixed sinusoidal electric field $E$, which helps to improve the signal-noise ratio, to avoid influence of any current shunting, and to minimize the impact of Joule heating. Here, $E$ of 50 kV/m is applied by the input voltage ($V_{\rm in}$) of 4 V onto the 60 μm long Hall bar (Fig. 1b), i.e., $E = 0.75V_{\rm in}/L$. The factor 0.75 arises from the lateral electric field spreading at the Hall crosses due to the presence of the detection leads (see Fig. 1c and below). No thermal annealing was performed for the devices in this work to avoid uncertainty from interface variations.

Microscopic distributions of the electric current and fields in the NM/FM bilayers are simulated using finite-element analysis using a very small unit cell size of 0.2 nm×0.2 nm×0.2 nm (similar to that in ref. [79]). The finite-element analysis follows the current continuity equation $\boldsymbol{\nabla}\cdot\boldsymbol{j_c}+\frac{\partial\rho}{\partial t}=0$ and the boundary condition of $\boldsymbol{n}\cdot\boldsymbol{j}_{\rm c}=0$, where $\boldsymbol{j}_{\rm c}$ is the charge current density, $\rho$ is the charge density, $\boldsymbol{n}$ is the normal direction of the device boundary. The analysis has taken into account the large Pt electrodes and the dimensions of the real device (Fig. 1d). The SOT of the representative Hall devices of magnetic heterostructures are characterized using harmonic Hall voltage analysis to verify the mechanism and potential impact of the scaling of $V_{\rm AH}$ with the NM.

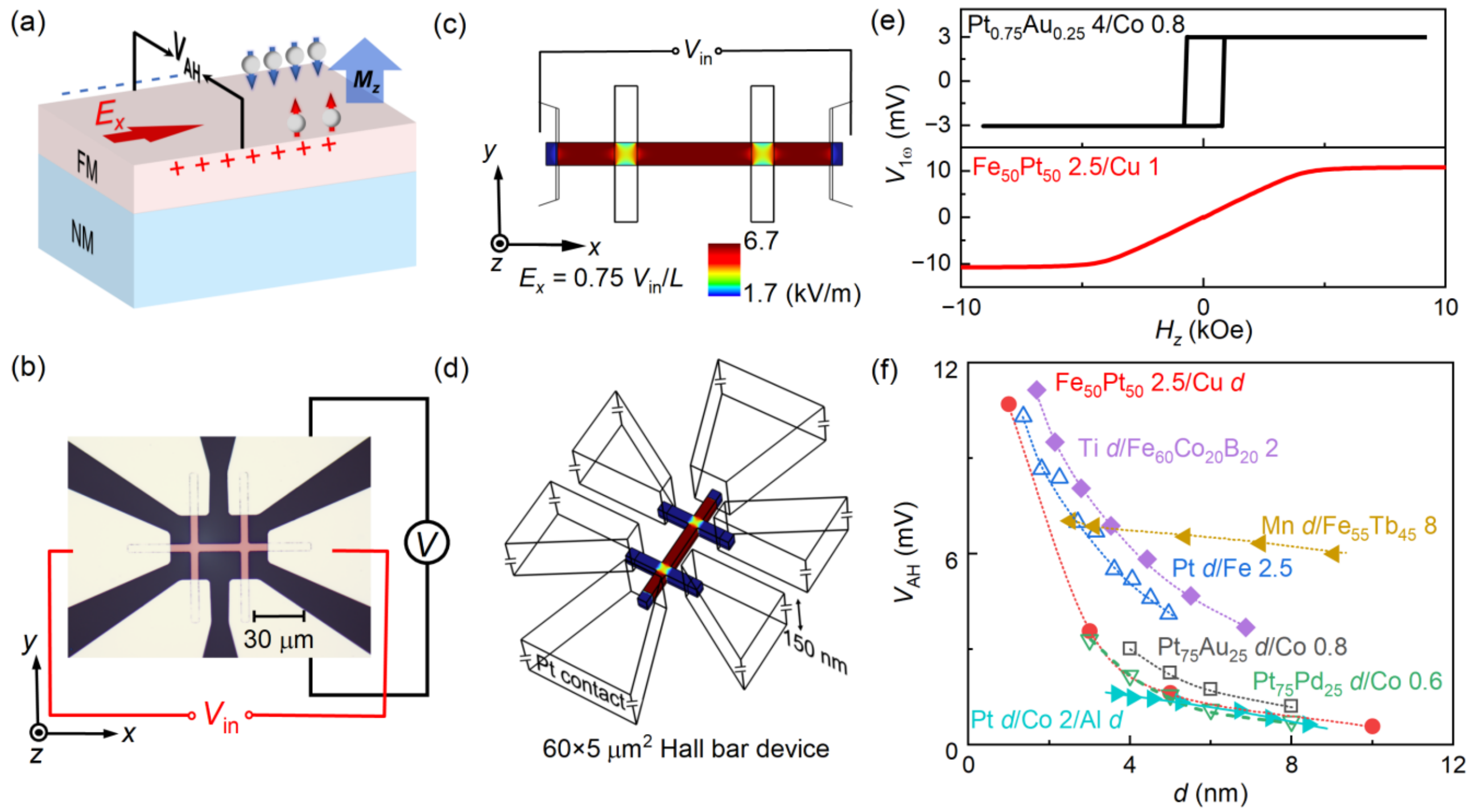

**Figure 1.** (a) Schematic diagram of the anomalous Hall effect. (b) Optical microscope image of the $Fe_{50}Pt_{50}$ 2.5/Cu 10 Hall bar device and measurement geometry of the anomalous Hall voltage. (c)The electric field applied to the cross region is $0.75V_{in}/L$ when a voltage $V_{in}$ is applied to the Hall bar with length $L$. (d) Device modeling and mesh division in the finite element analysis. (e) $V_{1\omega}$ *vs.* out-of-plane magnetic field $H_z$ for the $Pt_{75}Au_{25}$ 4/Co 0.8 and the $Fe_{50}Pt_{50}$ 2.5/Cu 1. (f) Dependence on normal metal thickness of the anomalous Hall voltage for $Fe_{50}Pt_{50}$ 2.5/Cu *d*, Pt 5/Co 2/Al *d*, $Pt_{75}Au_{25}$ *d*/Co 0.8, $Pt_{75}Pd_{25}$ *d*/Co 0.6, Pt *d*/Fe 2.5, Mn *d*/$Fe_{55}Tb_{45}$ 8, Ti *d*/$Fe_{60}Co_{20}B_{20}$ 2 heterostructures.

*Universal scaling of $V_{AH}$ with the normal metal*. The values of $V_{AH}$ for the $Fe_{50}Pt_{50}$ 2.5/Cu *d*, Pt 5/Co 2/Al *d*, $Pt_{75}Au_{25}$ *d*/Co 0.8, $Pt_{75}Pd_{25}$ *d*/Co 0.6, Pt *d*/Fe 2.5, Mn *d*/$Fe_{55}Tb_{45}$ 8, Ti *d*/$Fe_{60}Co_{20}B_{20}$ 2 are extracted from the dependence of $V_{1\omega}$ on the swept out-of-plane magnetic field ($H_z$) (Fig.1e). As shown in Fig. 1f, $V_{AH}$ of each sample series decreases very rapidly at small thicknesses and then more gradually upon further increasing the thickness. This observation suggests that it is a rather universal effect that $V_{AH}$ of magnetic heterostructures decreases upon increasing the thickness of the adjacent NM.

*Finite-element simulation of the $V_{AH}$ scaling*. The universal scaling of $V_{AH}$ with the adjacent normal metal suggests that there is an intrinsic mechanism that cause the anomalous Hall voltage of a magnetic layer to decrease with increasing the thickness of the adjacent NM layer. In general, there are two effects that could potentially affect the measured values of $V_{AH}$ of magnetic heterostructures, i.e., the non-uniform distribution of electric field (equivalently current) induced by the current spreading and the charge screening effect by the NM as a serial resistor.

As we model in the Fig.2a, when an excitation voltage $V_{in}$ is applied from the top surfaces of the thick, highly conductive current electrodes (with a thickness of 150 nm and a resistivity of 10 μΩ cm) into the thin magnetic heterostructures, there would be strong spatial spreading of the current and thus electric field in the vertical direction. Qualitatively, such vertical spreading creates non-uniformity in the current distributions in the NM and the magnetic layer, which further affects the experimentally measured values of $V_{AH}$. To evaluate this effect, we have performed finite element analysis of the FM/NM (Fig. 2a) and NM/FM bilayers with opposite order of the FM and the NM. As summarized in Fig. 2c, the simulated ratio of the current flow in the FM layer relative to the total current ($I_{FM}/I_{tot}$) for the FM 5/NM *d* and NM *d*/FM 5 bilayers ($\rho_{FM}/\rho_{NM}$ = 8) decreases quasi-linearly by less than 5% and 4 %, respectively, even when *d* is dramatically increased from 0 nm to 100 nm. For the FM 5/NM 5 and NM 5/FM 5 bilayers with fixed layer thicknesses, $I_{FM}/I_{tot}$ also decreases by less than 2.5% and 3.5%, respectively, as $\rho_{FM}/\rho_{NM}$ is increased from 0 to 100. These results suggest that the vertical current spreading effect only has negligible influence on the measurement of the anomalous Hall voltage of practical spintronic devices with $\rho_{FM}/\rho_{NM}$ of <10 and $t_{NM}$ < 10 nm.

The charge screening effect of the NM on the detection of the anomalous Hall voltage is schematically modeled in Fig. 3a. In general, the FM and the NM can be considered as two resistors in a serial. the FM generating the anomalous Hall voltage can be regarded as a battery with output $V_{AH,0}$, the two ends of which are connected to the NM and the voltmeter (Lock-in amplifier with internal resistance of 1 MΩ) at the same time. In this case, what the voltmeter measures is the voltage distributed onto the NM rather than $V_{AH,0}$ of the FM. In other words, the value of $V_{AH}$ experimentally measured from the NM is only a portion of that generated by the FM battery, while the voltage distribution in the FM is also significant. From the finite-element analysis (Fig. 3b), we simulated the Hall voltage between the two detection leads of a NM/FM Hall device ($\rho_{AH}$ = 50 μΩ cm, $\rho_{FM}$ = 1000 μΩ cm) under an electric field of $E$ = 50 kV/m by taking into account the voltage distributions on the FM and the NM layers. As summarized in Fig. 3c, the apparent values of $V_{AH}$ from the measurement, i.e., the value of $V_{AH}$ distributed on the NM, decreases dramatically towards vanishingly small value for the FM 2/NM $d$ ($\rho_{FM}/\rho_{NM}$ = 8) as $d$ increases and for the FM 2/NM 5 as $\rho_{FM}/\rho_{NM}$ increases, which is in excellent agreement with the experiments in Fig. 1f.

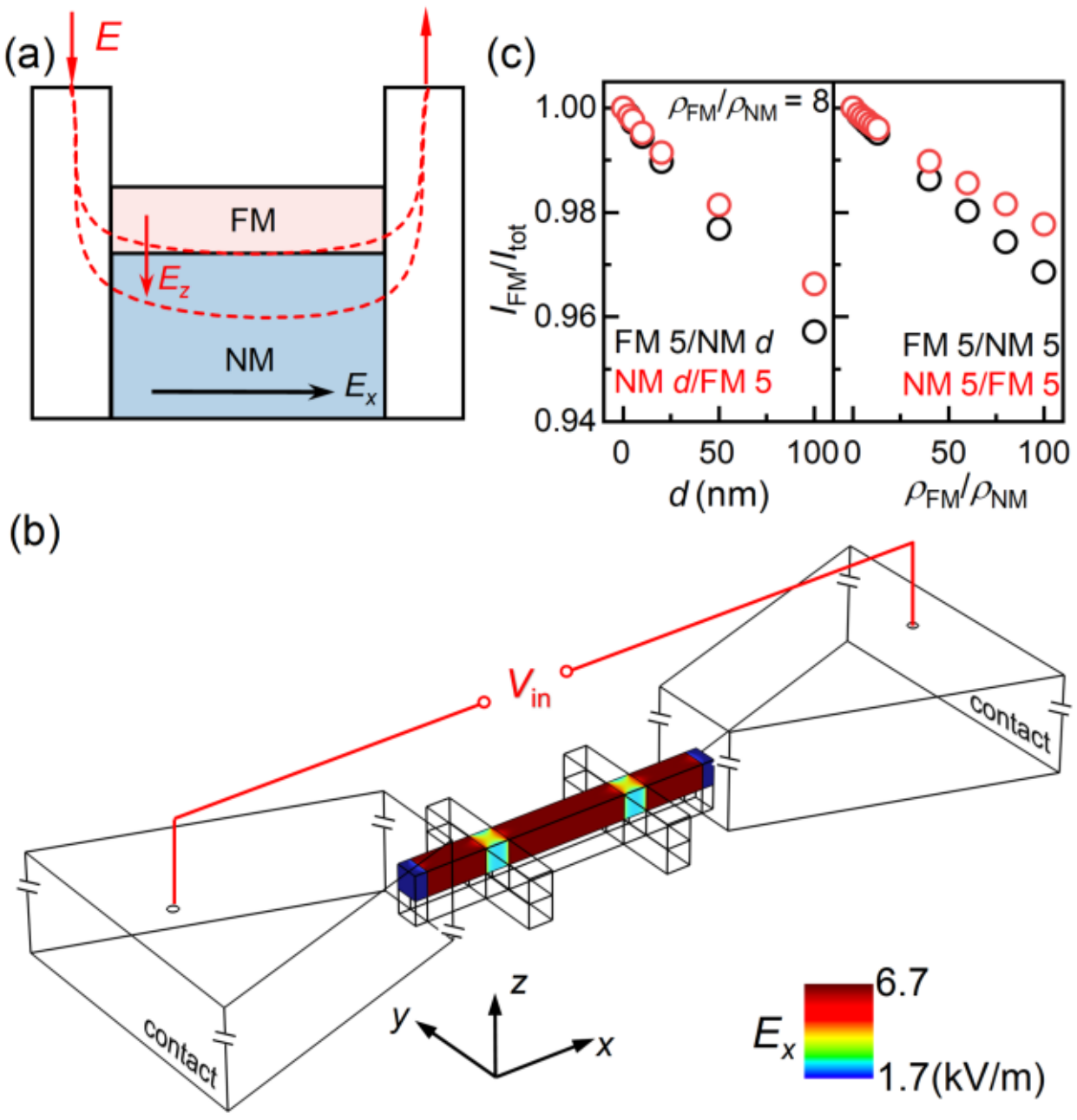

**Figure 2.** (a) Schematic and (b) Finite-element modeling of the electric field distributions in a NM/FM Hall bar with the electrical contact. (c) Simulated results of $I_{FM}/I_{tot}$ vs the thickness of the NM and the ratio of the resistivity of the FM and NM. Here, $I_{FM}/I_{tot}$ is the ratio of the volumetric integral of the longitudinal current density in the cross region of the FM in a bilayer Hall bar to that in a single-layer Hall bar with no adjacent conducting layer.

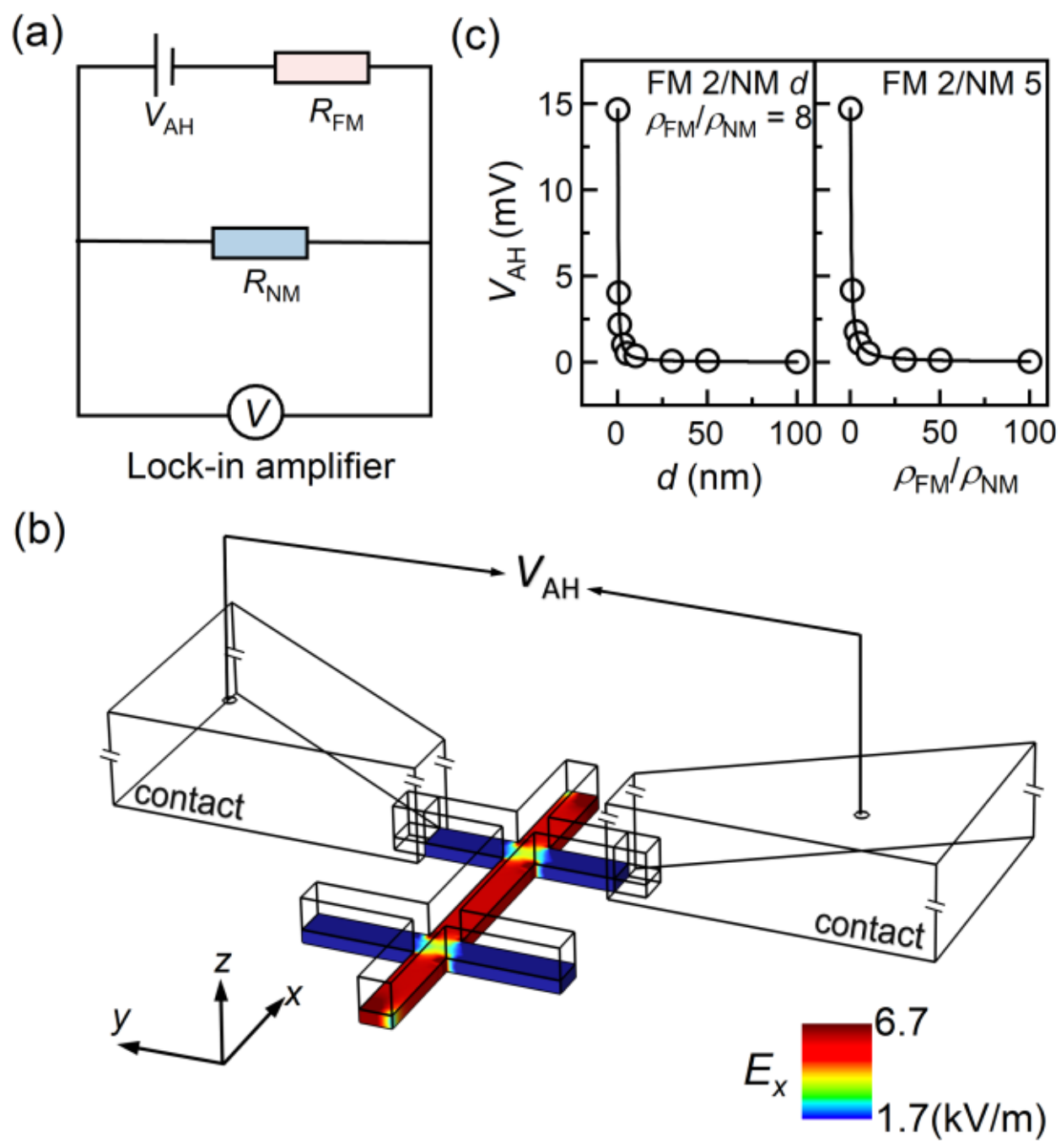

**Figure 3.** (a) Schematic and (b) Finite-element modeling of the charge screen effect of NM/FM heterostructure on the detection of the transverse voltage via the resistor serial. (c) Scaling of the simulated transverse voltage ($V_{AH}$) for FM 2/NM $d$ ($\rho_{FM}/\rho_{NM}$ = 8) and FM 2/NM 5 with the thickness and resistivity of the NM.

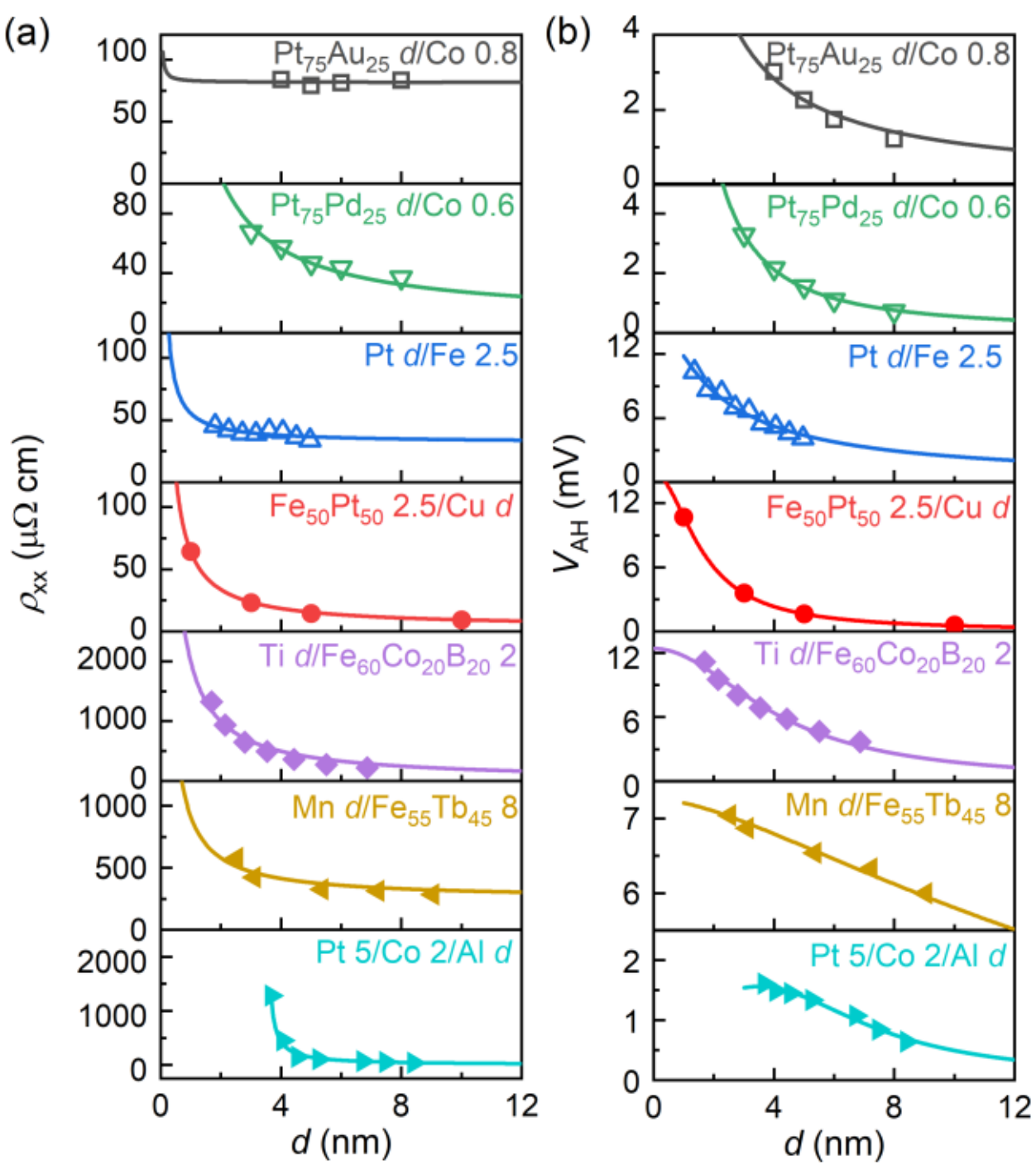

**Figure 4**. Dependence on the NM thickness of (a) the resistivity of normal metals and (b) the anomalous Hall voltage for the $Pt_{75}Au_{25}$ $d$/Co, $Pt_{75}Pd_{25}$ $d$/Co, Pt $d$/Fe, Ti $d$/$Fe_{60}Co_{20}B_{20}$, $Fe_{50}Pt_{50}$/Cu $d$, Mn $d$/$Fe_{55}Tb_{45}$ and Pt/Co/Al $d$ heterostructures.

*Analytical modeling of the $V_{AH}$ scaling*. Based on the screening effect of the NM via the serial resistor model, the experimentally measured $V_{AH}$ can be derived as

$$V_{AH} = V_{AH,0} /(1+ d\rho_{FM}/t\rho_{NM}). \qquad (2)$$

where $V_{AH,0}$ is the anomalous Hall voltage of the FM in the absence of an adjacent NM layer. The magnitude of $V_{AH,0}$ is independent of the parameters of the NM.

As indicated by the solid curves in Fig. 3c, Eq. (2) fits well the scaling of $V_{AH}$ with $d$ and $\rho_{FM}/\rho_{NM}$ from the finite-element analysis.

As shown in Fig.4a, for various practical NM $d$/FM $t$ bilayers with small $d$ of a few nanometers, $\rho_{NM}$ is not always constant but typically gets enhanced upon decreasing $d$ due to the interfacial scattering. Phenomenologically, the $d$ dependence of $\rho_{NM}$ usually follows

$$\rho_{NM} = \rho_0 + \rho_{int}/d, \qquad (3)$$

where $\rho_0$ is the thickness-dependent resistivity contribution, and $\rho_{int}$ parameterizes the interfacial scattering contribution when $d$ is comparable or smaller than the finite mean-free path of the NM. In Fig. 4a, the $d$ dependence of the average resistivity of the NM is fit well by Eq. (4), from which the values of $\rho_0$ and $\rho_{int}$ for various NM/FM bilayer are subtracted. Taking into account the interfacial scattering, Eq. (2) is rewritten as

$$V_{AH} = V_{AH,0} /(1+ d\rho_{FM}/t(\rho_0 + \rho_{int}/d)). \qquad (4)$$

As shown in Fig. 4b, Eq. (3) can fit strong dependence on $d$ of the measured anomalous Hall voltage for all the studied bilayers, regardless of the types of the FM and NM.

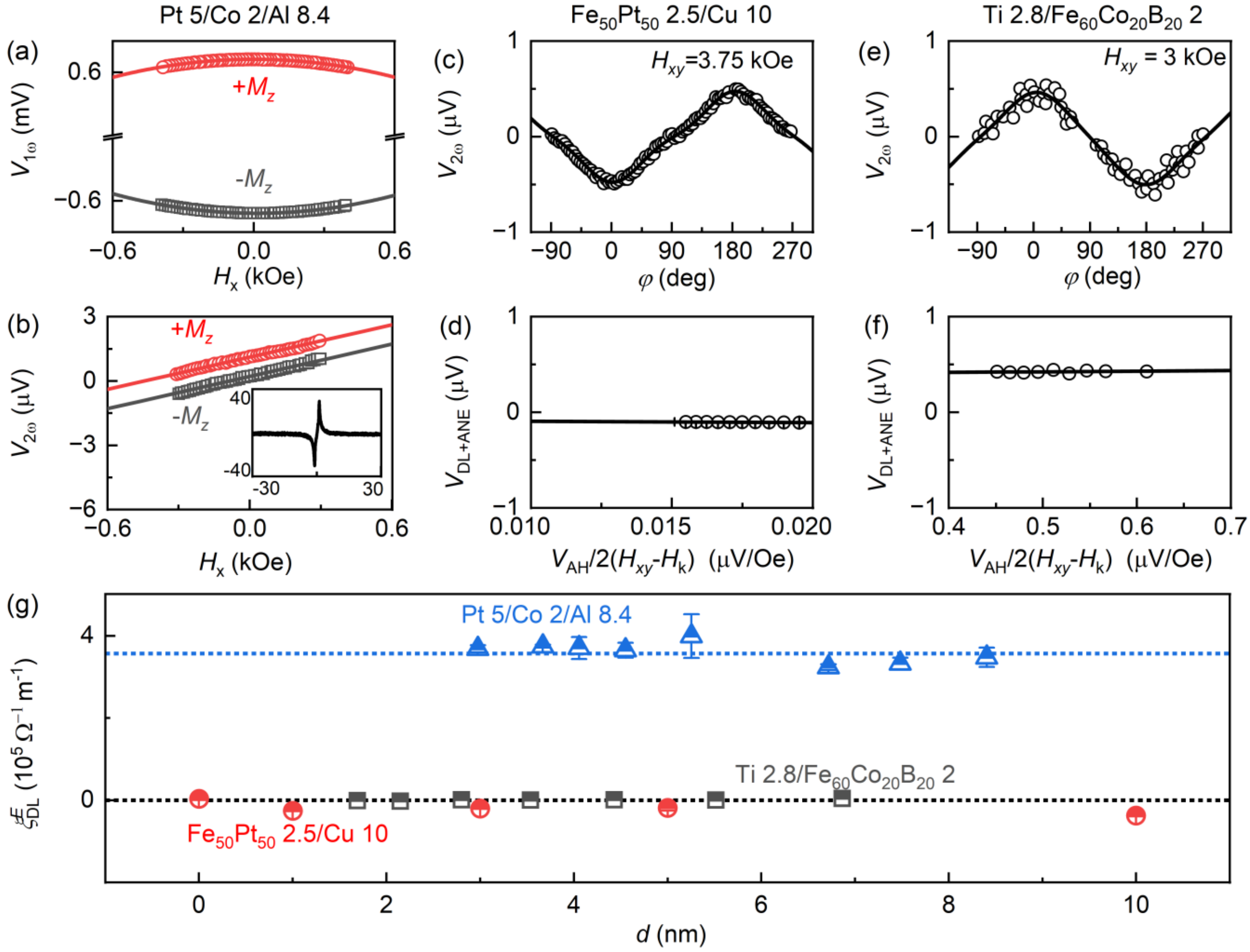

**Figure 5**. Dependence on $H_x$ of (a) First Harmonic Hall voltage and (b) Second harmonic Hall voltage for the Pt 5/Co 2/Al 8.4 with perpendicular magnetic anisotropy. (c) Second harmonic Hall voltage vs. azimuth angle ($\varphi$) of the in-plane magnetic field ($H_{xy}$ = 3.75 kOe) and (d) $V_{\mathrm{DL+ANE}}$ vs. $V_{\mathrm{AH}}/2(H_{xy}$-$H_{\mathrm{k}})$ for the $Fe_{50}Pt_{50}$ 2.5/Cu 10. (e) Second harmonic Hall voltage vs $\varphi$ ($H_{xy}$ = 3 kOe) and (f) $V_{\mathrm{DL+ANE}}$ vs. $V_{\mathrm{AH}}/2(H_{xy}$-$H_{\mathrm{k}})$ for the Ti 2.8/$Fe_{60}Co_{20}B_{20}$ 2. (g) Dampinglike SOT efficiency per electric field for the Pt/Co/Al $d$, $Fe_{50}Pt_{50}$/Cu $d$ and Ti $d$/$Fe_{60}Co_{20}B_{20}$ with different NM thickness.

*Irrelevance to angular momentum injection.* Given the previous speculation that the scaling of $V_{\mathrm{AH}}$ with the NM arisen from the Fermi level properties being altered by spin angular momentum injected from the NM [63], we verify that the strong $d$ dependence of the measured anomalous Hall voltage in NM/FM bilayers generally occurs in the absence of any spin angular momentum injection from the thickness-varying NM to the FM. We detect the angular momentum injection from the NM to the FM by the spin-orbit torque efficiency.

For this purpose, we perform harmonic Hall voltage analyses with $E$ = 0.75 × 66.7 kV/m to quantify the dampinglike SOT efficiencies of three representative device series, i.e., perpendicularly magnetized Pt 5/Co 2/Al $d$, in-plane magnetized $Fe_{50}Pt_{50}$ 2.5/Cu $d$, in-plane magnetized Ti $d$ /$Fe_{60}Co_{20}B_{20}$ 2. In these devices, the spin Hall effect is negligible in the thickness-varying light 3$d$ NM layers (i.e., the Al, Cu, Ti) such that the true values of the torque should remain independent of $d$.

For the PMA Pt 5/Co 2/ Al $d$ devices, both $V_{1\omega}$ and $V_{2\omega}$ (Fig. 5b) are collected as a function of the swept in-plane longitudinal magnetic field ($H_x$) along the current direction. As shown in Fig. 5a, $V_{1\omega}$ is a parabolic function of $H_x$ as expected from the tilting of a macrospin, i.e.,

$$V_{1\omega} = V_{\mathrm{AH}}\,(1 - H_x^2/H_k^2), \qquad (5)$$

where $H_{\mathrm{k}}$ is the perpendicular magnetic anisotropy field. In Fig. 5b, $V_{2\omega}$ scales linearly with the $H_x$. We then determine the damping-like SOT field ($H_{\mathrm{DL}}$) following [80]

$$H_{\mathrm{DL}} = -2\frac{\partial V_{2\omega}}{\partial H_x} \Big/ \frac{\partial^2 V_{1\omega}}{\partial^2 H_x} - 2H_{\mathrm{k}}\frac{V_{\mathrm{ANE}}}{V_{\mathrm{AH}}}, \qquad (6)$$

where $V_{\mathrm{ANE}}$ is the anomalous Nernst voltage induced by the vertical thermal gradient and equal to the value of $V_{2\omega}$ when magnetization is fully aligned to the current direction by $H_x$ that is much greater than $H_{\mathrm{k}}$ (see the inset of Fig. 5b).

For the in-plane magnetized $Fe_{50}Pt_{50}$ 2.5/Cu $d$ and Ti $d$ /$Fe_{60}Co_{20}B_{20}$ 2 devices, $V_{2\omega}$ is collected as a function of the azimuth angle ($\varphi$) of the in-plane magnetic field $H_{xy}$ relative to the current direction (Fig. 5b,c). $V_{2\omega}$ of an in-plane macrospin is given by [81]

$$V_{2\omega} = V_{\mathrm{DL+ANE}}\cos\varphi + V_{\mathrm{FL+Oe}}\cos\varphi\cos 2\varphi + V_{\mathrm{PNE}}\sin 2\varphi, \qquad (6)$$

$$\text{with } V_{\mathrm{DL+ANE}} = V_{\mathrm{AH}}H_{\mathrm{DL}}/2(H_{xy} - H_{\mathrm{k}}) + V_{\mathrm{ANE}}. \qquad (7)$$

Here, $V_{\mathrm{PNE}}$ is the planar Nernst voltage induced by the longitudinal thermal gradient. The fit of the $\varphi$ dependence of $V_{2\omega}$ to Eq. (6) yields the values of $V_{\mathrm{DL+ANE}}$ and $V_{\mathrm{FL+Oe}}$ for each magnitude of $H_{xy}$. As shown in Fig. 5d,f, the $H_{\mathrm{DL}}$ values of the in-plane magnetized devices are determined from the slope of the linear fits of $V_{DL+ANE}$ vs $V_{AH}/2(H_{xy} - H_k)$ following Eq. (7). With the values of $H_{DL}$, the dampinglike SOT efficiency $\xi_{DL}^{E}$ is calculated following

$$\xi_{DL}^{E} = (2e/\hbar)\mu_0 M_s t H_{DL}/E, \quad (8)$$

where $e$ is the elementary charge, $\hbar$ the reduced Planck constant, $\mu_0$ the permeability of vacuum. As shown in the Fig. 5g, for all the device series, $\xi_{DL}^{E}$ remains essentially constant despite the strong variation of $V_{AH}$ with the NM thickness $d$. These results unambiguously reveal that the strong, universal scaling of the measured anomalous Hall voltage with the adjacent NM thickness $d$ is irreverent to any modification of the Fermi surface properties or angular momentum injection from the NM to the FM. Here, we note that the independence of $\xi_{DL}^{E}$ on the thickness of the NM is consistent with the negligible orbital torque contributions from the Al, Ti, and Cu in previous reports [30,74-76,78].

*Robustness of the harmonic Hall voltage analysis*. Since the damping-like SOT is identified from its interplay with the anomalous Hall voltage in the harmonic Hall voltage analyses, we discuss that the misreading of $V_{AH}$ due to the into serial resistor effect of the NM does not affect the strength of the SOTs from the harmonic Hall voltage analysis. According due to serial resistor effect of the NM/FM bilayer and Eq. (2), any transverse voltages are generally lowered by the same factor of $1/(1+ d\rho_{FM}/t\rho_{NM})$ compared to their true values. Thus, all the voltages in Eqs. (5)-(7), $V_{1\omega}$, $V_{2\omega}$, $V_{ANE}$, $V_{DL+ANE}$, and $V_{AH}$, are lowered by the same factor, leading to no alteration to the values of $H_{DL}$ and the torque efficiency of PMA and in-plane magnetic devices as can be seen from Eq. (5) and Eq. (7).

*Conclusion.* We have presented a systematic study on the observation, mechanism, and impact of the universal scaling of the anomalous Hall voltages in magnetic heterostructures with varying NM thicknesses by combining transport experiments, analytical derivation, and finite-element analyses. We show that the anomalous Hall voltage reduces universally and dramatically with the presence and thickness of the adjacent non-magnetic conducting layer, following Eq. (2). The dominant mechanism is the serial resistor effect of the magnetic and non-magnetic layers within the magnetic heterostructures. We have also verified that such scaling of the measured anomalous Hall voltages is irrelevant to any Fermi surface variation or any angular momentum injection from the nonmagnetic layer to the magnetic layer. We have also discussed that the accuracy of the harmonic Hall voltage analyses of PMA and in-plane magnetized magnetic heterostructures are unaffected by the under-measuring of the anomalous Hall voltage and other transverse voltages due to the serial resistor effect. These findings provide crucial information for understanding a variety of spintronic phenomena involving the anomalous Hall and other transverse voltages.

This work is supported partly by the National Key Research and Development Program of China (2022YFA1204000), the Beijing Natural Science Foundation (Z230006), and the National Natural Science Foundation of China (12274405, 12304155)